\documentclass[conference]{IEEEtran}
\IEEEoverridecommandlockouts

\usepackage{cite}
\usepackage{amsmath,amssymb,amsfonts}
\usepackage{algorithmic}
\usepackage{graphicx}
\usepackage{textcomp}
\usepackage{xcolor}
\usepackage{times}
\usepackage{graphicx}
\usepackage{booktabs}
\usepackage{url}
\usepackage{tikz}
\usepackage{balance}
\usepackage{caption}
\usepackage{pdfpages}
\usetikzlibrary{shapes.geometric, arrows.meta, positioning}

\tikzstyle{process} = [rectangle, minimum width=1.0cm, minimum height=0.6cm, text centered, draw=black, fill=blue!10]
\tikzstyle{io} = [trapezium, trapezium left angle=70, trapezium right angle=110, minimum height=0.6cm, text centered, draw=black, fill=orange!20]
\tikzstyle{arrow} = [thick, ->, >=stealth]

\usepackage{pifont}
\definecolor{cmarkcolor}{RGB}{21, 164, 64} 
\definecolor{xmarkcolor}{RGB}{177, 0, 4} 

\newcommand{\cmark}{\color{cmarkcolor}\ding{51}}%
\newcommand{\xmark}{\color{xmarkcolor}\ding{55}}%

\title{Assessing and Enhancing Quantum Readiness in Mobile Apps}

\author{\IEEEauthorblockN{Joseph Strauss\IEEEauthorrefmark{1}, Krishna Upadhyay\IEEEauthorrefmark{1}, A.B. Siddique\IEEEauthorrefmark{2}, Ibrahim Baggili\IEEEauthorrefmark{1}, Umar Farooq\IEEEauthorrefmark{1}}
\IEEEauthorblockA{\IEEEauthorrefmark{1}
    Louisiana State University, 
    \IEEEauthorrefmark{2} 
    University of Kentucky \\
 Email: 
jstrau9@lsu.edu, kupadh4@lsu.edu, siddique@cs.uky.edu, ibaggili@lsu.edu, ufarooq@lsu.edu
}}

\begin{document}

\maketitle

\begin{abstract}
Quantum computers threaten widely deployed cryptographic primitives such as RSA, DSA, and ECC. 
While NIST has released post-quantum cryptographic (PQC) standards (e.g., Kyber, Dilithium), mobile app ecosystems remain largely unprepared for this transition.
We present a large-scale binary analysis of over 4,000 Android apps  to assess cryptographic readiness. 
Our results show widespread reliance on quantum-vulnerable algorithms such as MD5, SHA-1, and RSA, while PQC adoption remains absent in production apps. 
To bridge the readiness gap, we explore LLM-assisted migration.
We evaluate leading LLMs (GPT-4o, Gemini Flash, Claude Sonnet, etc.) for automated cryptographic migration. 
All models successfully performed simple hash replacements (e.g., SHA-1 to SHA-256). 
However, none produced correct PQC upgrades due to multi-file changes, missing imports, and lack of context awareness. 
These results underscore the need for structured guidance and system-aware tooling for post-quantum migration.
\end{abstract}

\section{Introduction}
Quantum computing threatens to render widely deployed cryptographic algorithms obsolete. Once sufficiently large quantum computers become available, public-key schemes such as RSA, Diffie-Hellman, and elliptic curve cryptography (ECC) will no longer provide meaningful security guarantees~\cite{nist-pqc-report}. 
In response to this threat, the U.S. National Institute of Standards and Technology (NIST) has selected several post-quantum cryptographic (PQC) algorithms for standardization, including CRYSTALS-Kyber for key encapsulation and CRYSTALS-Dilithium for digital signatures~\cite{nist-pqc}. 
While these standards mark a critical step forward in cryptographic resilience, there remains an open question as to whether mobile software ecosystems are prepared for this transition.

In this work, we present a comprehensive static analysis of Android applications to assess their quantum readiness. We analyze 4,018 apps from diverse categories, applying a lightweight static analysis capable of extracting cryptographic usage from Android apps. 
In addition to measuring the prevalence of quantum-vulnerable algorithms, we explore whether current development practices show any early adoption of post-quantum standards. We further investigate the role of large language models (LLMs) in assisting developers with cryptographic upgrades, particularly in transitioning from legacy algorithms to quantum-safe alternatives.

\section{Threat Model}
\begin{figure}[h]
    \centering
    \includegraphics[width=0.96\linewidth]{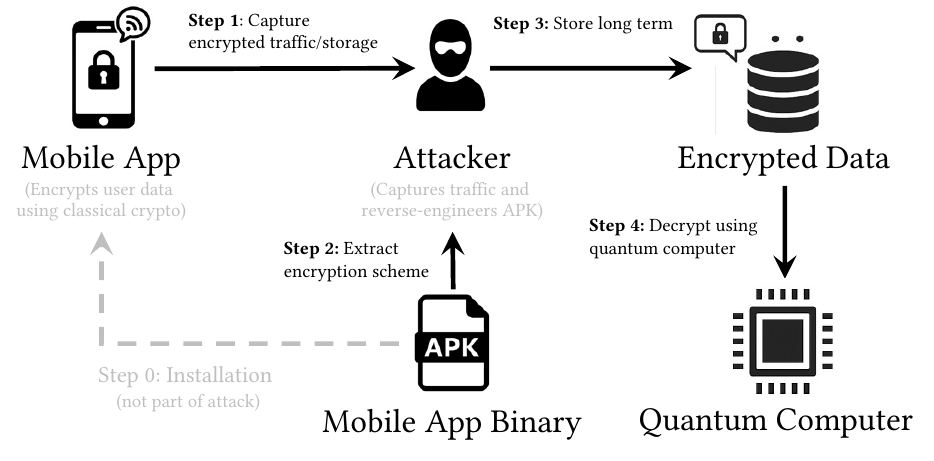}
    \vspace{-5pt}
\caption{Harvest Now, Decrypt Later (HNDL) attack model where encrypted mobile app data is captured and stored for future decryption using quantum computers.}
\label{fig:hndl-attack}
\vspace{-20pt}
\end{figure}

This work considers an adversary with access to a large-scale, fault-tolerant quantum computer capable of executing Shor's algorithm~\cite{shor1999polynomial} to break classical public-key cryptographic systems. 
The threat model assumes that any data encrypted or signed using RSA, DSA, or ECC can be compromised once quantum computers reach sufficient scale. 
As Fig~\ref{fig:hndl-attack} presents, the adversary is assumed to be capable of conducting ``harvest now, decrypt later'' attacks, where encrypted data -- collected today using classical schemes -- is stored for future decryption once quantum capabilities become available.

The adversary may also inspect or reverse engineer executable files to discover vulnerable cryptographic usages, taking advantage of weak hash functions (e.g., MD5, SHA-1), insecure cipher modes (e.g., ECB), or misuse of APIs (e.g., non-random IVs or insecure random number generation). 
The model does not assume zero-day vulnerabilities in the Android operating system or privileged access to the device; rather, it focuses on the cryptographic surface exposed by mobile apps.

The primary assets at risk include the confidentiality and long-term integrity of sensitive user data -- such as authentication tokens, messages, financial records, and health information -- as well as the security of communication channels and update mechanisms. 
The goals of our analysis and migration system are to identify instances of quantum-vulnerable cryptographic primitives in Android apps, assess the scale and pervasiveness of these vulnerabilities, and evaluate whether automated tools, including large language models, can assist in migrating apps to quantum-safe cryptographic standards.

This threat model motivates both the measurement and mitigation aspects of our work. By characterizing the potential impact of quantum-capable adversaries, we provide a framework for evaluating the necessity and feasibility of post-quantum cryptographic adoption in mobile ecosystems.

\section{Methodology}

\begin{figure}[t]
    \centering
    \includegraphics[width=0.96\linewidth]{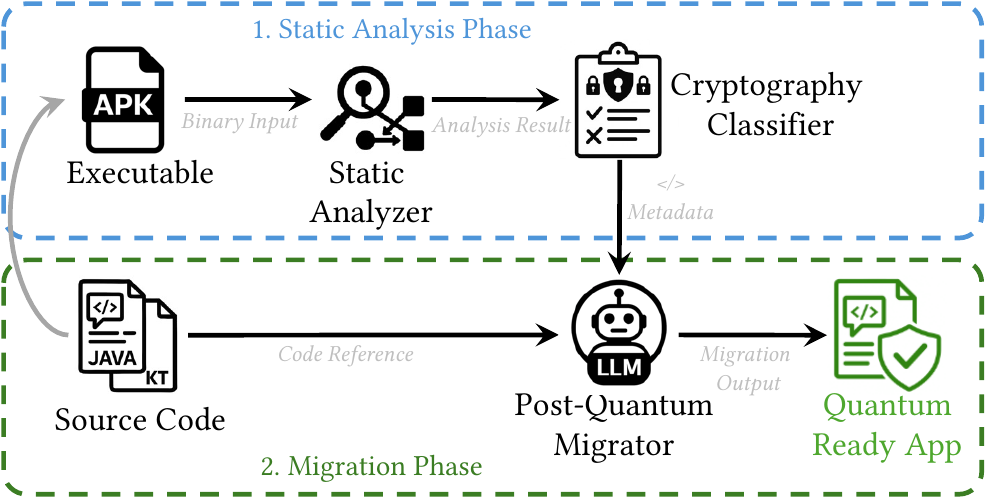}
    \vspace{-7pt}
\caption{System overview: Our two-phase system analyzes cryptographic usage in Android binaries and migrates legacy code to post-quantum cryptographic algorithms using LLMs.}
\label{fig:system-overview}
\vspace{-20pt}
\end{figure}
We perform static analysis of 4,018 Android apps collected from Google Play and F-Droid to assess their cryptographic readiness against quantum threats. 
We use CryptoAPI-Bench~\cite{CryptoAPI-Bench} inspired rules to detect cryptographic API usage in apps.
We perform backward dataflow analysis to resolve string arguments in factory methods such as \texttt{Cipher.getInstance()} and \texttt{KeyPairGenerator.getInstance()}, thereby recovering the exact algorithm used (e.g., \texttt{AES/CBC/PKCS5Padding}, \texttt{RSA/ECB/PKCS1Padding}).

Each detected instance is labeled as quantum-safe or quantum-vulnerable based on the underlying cryptographic algorithm. 
For example, AES and SHA-256 are considered safe (with caveats), while RSA, SHA-1, and MD5 are known to be broken by quantum algorithms~\cite{nist-pqc-report}.

Our pipeline, illustrated in Fig~\ref{fig:system-overview}, consists of two main phases. In Phase 1, static analysis extracts cryptographic usage patterns and passes metadata to a risk classifier. 
In Phase 2, we test whether LLMs can perform cryptographic upgrades using this metadata and corresponding code snippets. 
We evaluate state-of-the-art LLMs, including GPT-4o, Sonnet 3.7, Gemini Flash 2.0, and DeepSeek. 
We run each model in both edit mode and agentic mode (where supported).

Migration tasks range from hash upgrades (e.g., SHA-1 to SHA-256) to integration of post-quantum cryptographic algorithms (e.g., Kyber, Dilithium). 
The prompts are designed to reflect realistic developer workflows and include source files, target APIs, and sometimes PR-like diffs to simulate the reasoning a developer might perform.
We analyze whether these models can reason over multi-file contexts and modify usage patterns beyond local string replacements.
This enables an end-to-end analysis from cryptographic detection to automated remediation

\section{Results and Observations}
\begin{table}[t]
\centering
\caption{Most-used cryptographic algorithms in Android apps with quantum safety labels.}
\vspace{-8pt}
\label{tab:top-algos}
\begin{tabular}{lrrc}
\toprule
Algorithm & \# of instances & \# of Apps & Post-Quantum-Safe \\
\midrule
MD5 & 28,994 & 2,531 & {\xmark} \\
SHA-256 & 22,110 & 3,293 & \cmark  \\
SHA-1 & 18,200 & 2,454 & \xmark  \\
AES/CBC & 10,219 & 2,071 & {\cmark} $^\ast$  \\
RSA & 781 & 781 & {\xmark} \\
\bottomrule
\multicolumn{4}{l}{
       $^\ast$ Secure with 256-bit keys.}
\end{tabular}
\vspace{-20pt}
\end{table}

Our analysis reveals that mobile apps remain heavily reliant on quantum-vulnerable cryptographic primitives. As shown in Table~\ref{tab:top-algos}, SHA-256 is the most frequently used algorithm (3,293 apps), followed by SHA-1 (2,454 apps), MD5 (2,531 apps), and RSA (589 apps). AES is widely used, particularly in CBC mode with padding, but security varies based on configuration and key size. The presence of insecure or outdated algorithms suggests a systemic cryptographic debt across mobile apps.

Despite recent NIST standardization of PQC schemes such as Kyber and Dilithium, we found no evidence of PQC adoption in the analyzed APKs. 
F-Droid apps -- often considered more security-conscious -- do not use PQC in practice, although some apps include unused class-level imports referencing PQC libraries.

In LLM migration experiments, all tested models performed well on hash upgrades such as SHA-1 to SHA-256. These tasks involved minor changes to method arguments and were fully supported by existing APIs. 
For PQC integration, we collected GitHub repositories as reference examples. 
This migration requires adding new methods, updating multiple files, managing dependencies (e.g., Bouncy Castle PQC), and compatibility steps that were consistently omitted or incorrectly handled.
In our experiments, no model completed post-quantum migrations. 
Models like GPT-4o and Sonnet 3.7, even in agentic mode, produced partially correct suggestions but failed to generate secure or compilable patches. 
Placeholder function calls were sometimes inserted without correct imports or logic. 
These results highlight a critical gap: while LLMs are competent at local refactoring, they cannot handle security-critical migrations such as those required for PQC.

\section{Conclusion and Future Directions}
We present the first large-scale analysis of quantum readiness in Android apps. Our results show widespread use of quantum-vulnerable algorithms and no evidence of post-quantum adoption in production apps. 
While LLMs succeed at basic refactoring but fall short on secure, compilable PQC migrations.
This highlights the need for migration tools that understand dependencies, context, and security goals. 
Future work includes building a larger dataset for PRs, improved prompting, and integrating LLM-assisted crypto upgrades into developer workflows.

\balance
\bibliographystyle{plain}

\bibliography{IEEEexample.bib}
\end{document}